\documentclass[11pt, logo, twocolumn, copyright]{a2z}

\usepackage[authoryear, sort&compress, round]{natbib}
\usepackage{floatrow}
\usepackage{varwidth}
\usepackage[]{algorithm2e}
\usepackage{algpseudocode}
\usepackage{bbm}
\usepackage{wrapfig}
\usepackage{siunitx}
\usepackage{array}

\title{a2z-1 for Multi-Disease Detection in Abdomen-Pelvis CT: External Validation and Performance Analysis Across 21 Conditions}

\correspondingauthor{pranav@a2zradiology.ai}

\reportnumber{001} %

\author{
{\Authfont
    Pranav Rajpurkar PhD,
    Julian N. Acosta MD,
    Siddhant Dogra MD,
    Jaehwan Jeong MS,
    Deepanshu Jindal BS
    Michael Moritz MD,
    Samir Rajpurkar MS
\leavevmode \protect\\
{\Affilfont
\href{https://a2zradiology.ai/}{a2z Radiology AI}
}}}

\begin{abstract}
We present a comprehensive evaluation of a2z-1, an artificial intelligence (AI) model designed to analyze abdomen-pelvis CT scans for 21 time-sensitive and actionable findings. Our study focuses on rigorous assessment of the model's performance and generalizability. Large-scale retrospective analysis demonstrates an average AUC of 0.931 across 21 conditions. External validation across two distinct health systems confirms consistent performance (AUC 0.923), establishing generalizability to different evaluation scenarios, with notable performance in critical findings such as small bowel obstruction (AUC 0.958) and acute pancreatitis (AUC 0.961). Subgroup analysis shows consistent accuracy across patient sex, age groups, and varied imaging protocols, including different slice thicknesses and contrast administration types. Comparison of high-confidence model outputs to radiologist reports reveals instances where a2z-1 identified overlooked findings, suggesting potential for quality assurance applications.
\end{abstract}

\begin{document}
\maketitle

\section{Introduction}
Image interpretation is a cornerstone for medicine, offering essential information for diagnosing and managing a wide range of conditions. Among imaging methods, computed tomography (CT) is especially valuable due to its detailed cross-sectional views of the body's internal structures. Abdomen-pelvis CT scans are particularly useful for identifying time-sensitive conditions like infections, blockages, or internal bleeding which require immediate medical attention, but can be difficult to interpret due to diversity in anatomy and potential pathology. The challenges in abdomen-pelvis CT interpretation, along with the increasing volume of imaging studies and a shortage of radiologists, underscore the urgent need for innovative solutions \citep{liu_extra_2022, peng_radiologist_2022, smith2019trends}. Advances in AI have positioned it as an exceptional tool to address these complex, data-intensive tasks, offering radiologists powerful support in managing growing workloads with greater accuracy and speed \citep{rajpurkar_current_2023}. To address this, we present \textbf{a2z-1}, an advanced AI model designed to assist in the accurate analysis of abdomen-pelvis CT scans by detecting 21 conditions.

\begin{figure*}[t!]
    \centering
    \includegraphics[width=\linewidth]{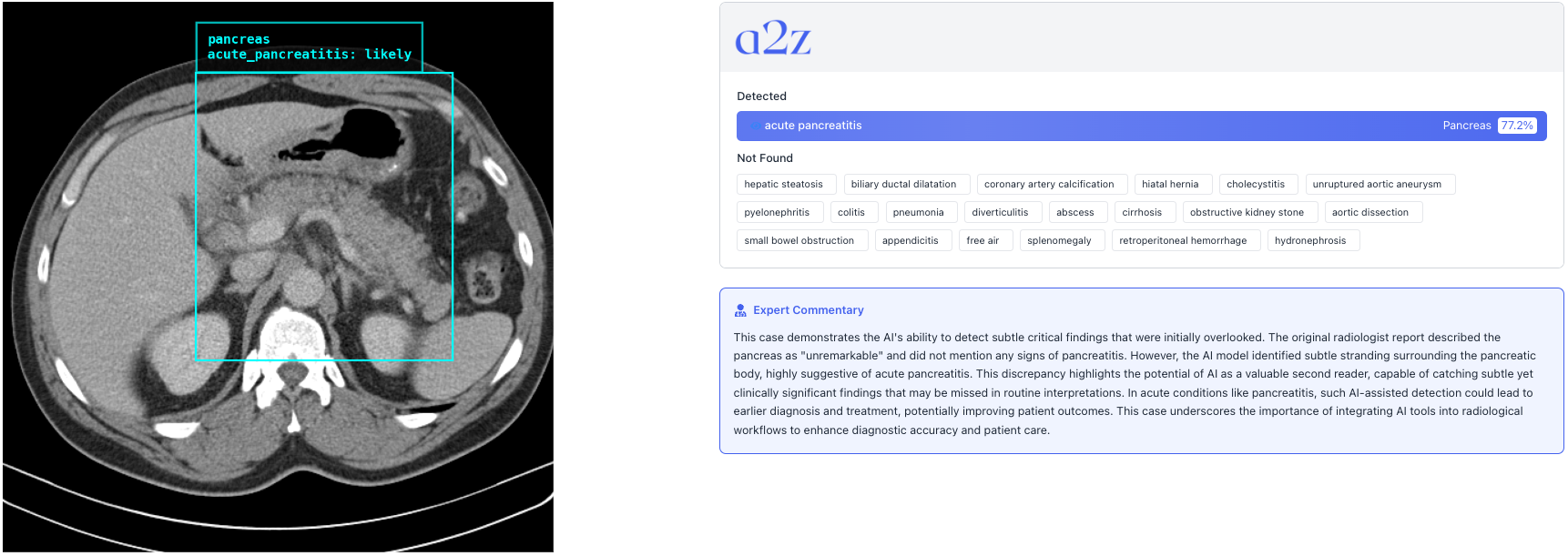}
    \caption{\textbf{a2z-1 detection of subtle pancreatitis overlooked in initial report.}}
    \label{fig:pancreatitis}
\end{figure*}

\paragraph{Challenges in CT Interpretation.} The interpretation of medical images, particularly CT scans, is an extraordinarily complex task that challenges even the most experienced radiologists \citep{itri2018fundamentals}. This complexity is especially pronounced in abdomen-pelvis CT interpretation, where radiologists must simultaneously analyze multiple intricate anatomical structures and identify a diverse array of potential conditions. Despite the high level of expertise in the field, the challenging nature of this task inevitably leads to variability in interpretations and, at times, errors. Studies have shown that up to 14\% of abdomen-pelvis CT reports undergo clinically important changes upon second review \citep{lauritzen2016radiologist}. Furthermore, significant disagreements can occur both between different radiologists examining the same CTs and when the same radiologist reviews an CTs at different times. These disagreement rates can be as high as 26\% and 32\% respectively \citep{abujudeh2010abdominal}, highlighting the subjective nature of image interpretation. A significant portion of radiological errors, estimated at 42\%, occur when abnormalities are simply not noticed or recognized during the initial review of an image \citep{kim2014fool}. These are known as perceptual errors, where the abnormality is present in the image but not consciously processed by the radiologist. Notably, such missed findings, rather than misinterpretations of observed abnormalities, are the primary reason for corrections in abdomen-pelvis CT reports \citep{rosenkrantz2016diagnostic, brady2017error}, highlighting an opportunity for AI-assisted reading to serve as a vigilant second observer \citep{rueckel_reduction_2021}.

\paragraph{AI's Potential in Abdomen-Pelvis CT.}
AI has emerged as a promising tool to address the unique challenges of abdomen-pelvis CT interpretation, with the potential to serve as a powerful assistant for radiologists. By aiding in the detection of pathologies in multiple organs, adapting to various types of study (e.g. contrast and non-contrast), identifying relevant incidental findings and being robust to the high variability of normal anatomy, AI could significantly enhance diagnostic accuracy and efficiency in abdominal imaging \citep{langlotz2023future}. However, the development of AI for abdomen-pelvis CT interpretation presents a unique and largely untapped opportunity. Only a few narrowly focused applications are currently available \citep{mervak_review_2023, burns_automated_2016, zhou_automatic_2021, chen_pancreatic_2023, cao_large-scale_2023, yin_value_2023}, creating a stark disparity between the complex reality of abdomen-pelvis CTs and the limited scope of existing AI models \citep{winkel_evaluation_2019, summers_atherosclerotic_2021}. This gap underscores a critical need for more comprehensive solutions.  To be truly effective and clinically valuable, these models must demonstrate excellent performance across a broad spectrum of conditions relevant to abdomen-pelvis CT, not just in isolated tasks. Moreover, these models must maintain their high performance when applied across different institutions and across various patient subgroups, ensuring equitable and consistent results regardless of factors such as age, sex, or imaging equipment.

\paragraph{Key Findings}
In response to the challenges and opportunities in abdomen-pelvis CT interpretation, we developed a2z-1, a novel AI model designed for multi-disease analysis of abdomen-pelvis CT scans. a2z-1 is capable of detecting and classifying 21 time-sensitive and actionable findings, representing a significant advancement in the application of AI to this complex domain. Our study provides a rigorous and comprehensive evaluation of a2z-1, with the following key findings:

\begin{figure*}[t!]
    \centering
\includegraphics[width=0.8\linewidth]{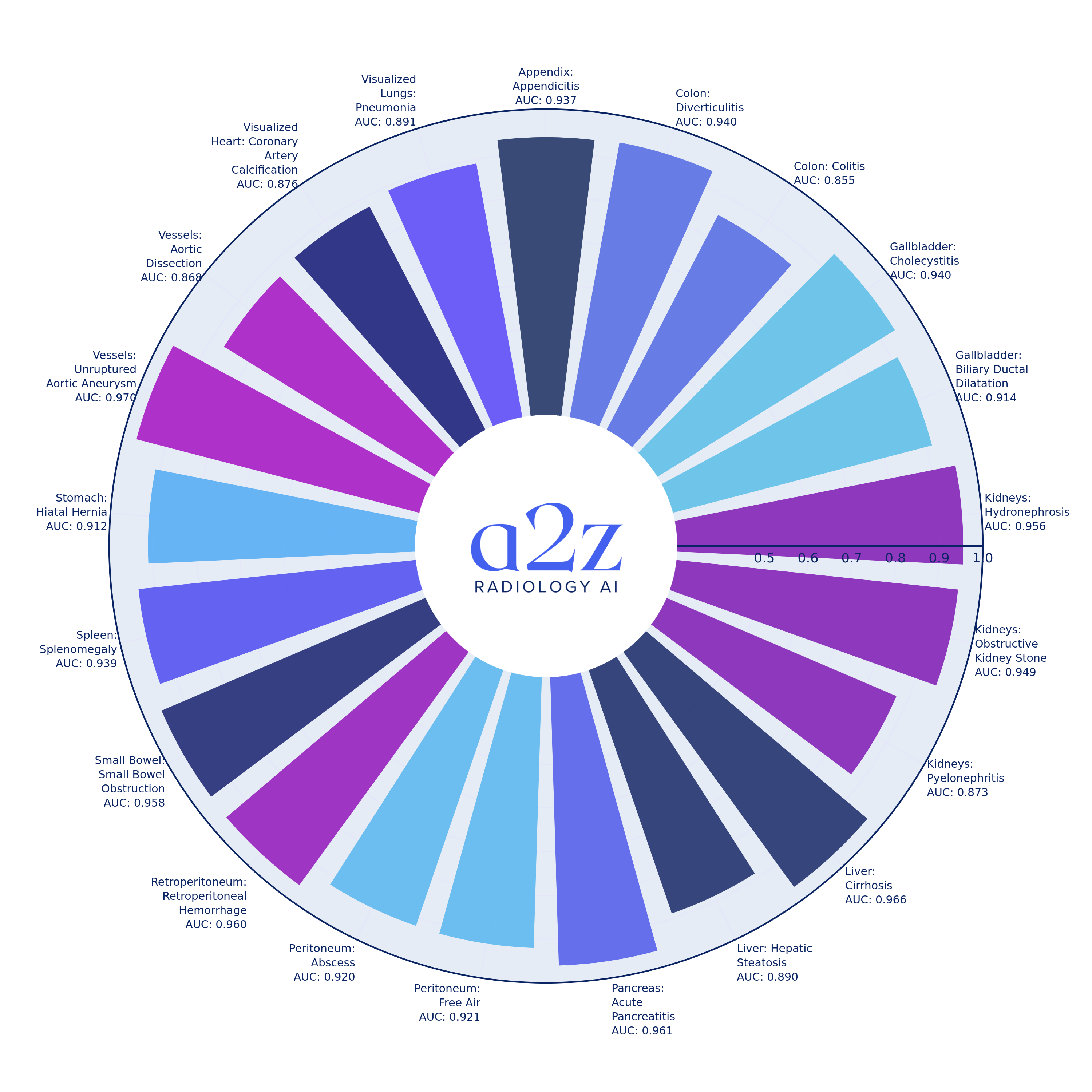}
\caption{\textbf{a2z-1 model performance across 21 abdominal conditions.} Each segment represents the model's AUC score for detecting a specific condition in abdominen-pelvis CT scans, demonstrating high and consistent performance across both internal and external validation datasets.}
    \label{fig:radial-plot}
\end{figure*}

\begin{itemize}
\item \textbf{Exceptional Performance in Detecting Critical Findings:} a2z-1 was evaluated across 21 key findings in abdomen-pelvis CT scans, achieving high performance with an average AUC of 0.923 on external validation. The model delivered high performance for certain time-sensitive conditions such as small bowel obstruction (AUC 0.958) and acute pancreatitis (AUC 0.961).
\item \textbf{Consistent Results Across Evaluation Scenarios and Patient Subgroups:} a2z-1 was assessed across three institutions, including two external sites not used during training. The model maintained strong performance across various patient groups, imaging protocols, and test conditions.
\item \textbf{Identification of Missed Findings and Workflow Optimization:} The a2z-1 model demonstrated its potential as a valuable second reader by identifying significant findings that were missed in initial radiologist reports. In a subset of cases, a2z-1 identified several clinically significant findings that were initially overlooked but were confirmed present by our expert review.
\end{itemize}

\section{Results}

\subsection{High Performance Across Diverse Conditions}

\paragraph{Selection of Conditions.} The a2z-1 model has been designed to detect a broad spectrum of abdominal conditions, with its performance assessed through Area Under the Curve (AUC) scores. The pathologies evaluated were chosen based on their clinical significance, guided by three main factors: urgency, potential for intervention, and prevalence. This approach ensures that the model is focused on conditions that demand timely treatment, have actionable outcomes, and are commonly encountered in clinical practice, allowing for robust performance measurement.

\begin{figure*}[t!]
    \centering
    \includegraphics[width=\linewidth]{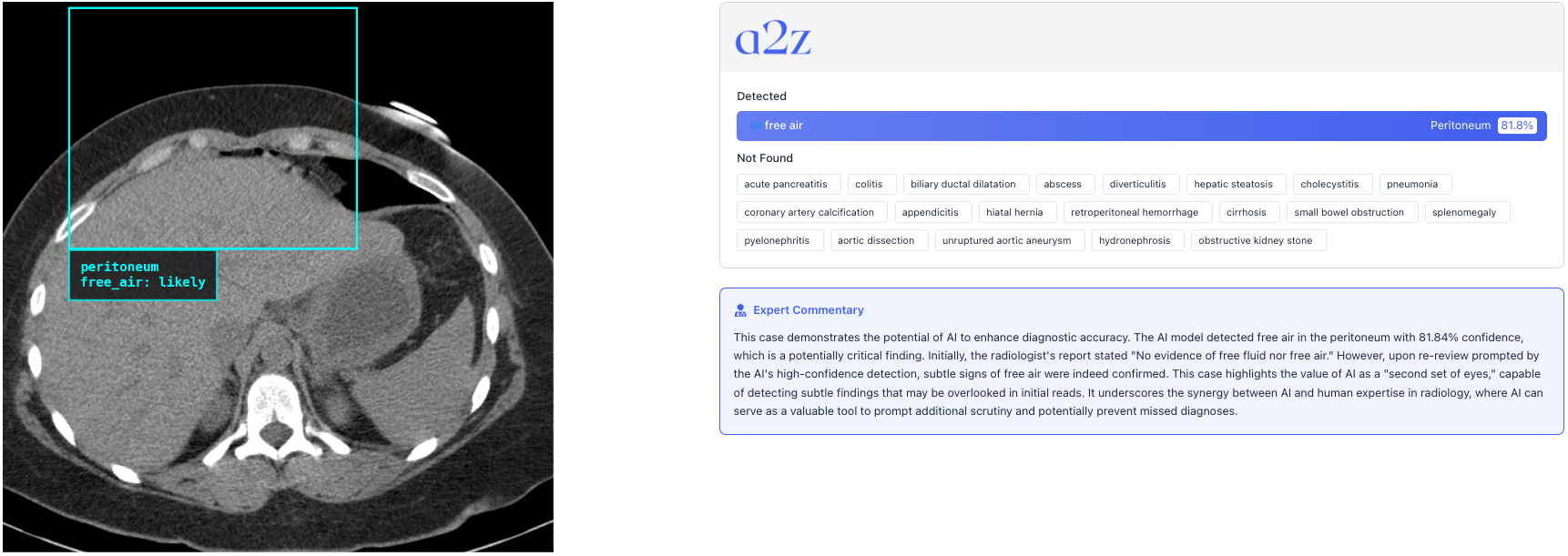}
    \caption{\textbf{a2z-1 detects free air finding missed in initial read.}}
    \label{fig:free-air}
\end{figure*}

\paragraph{AUC Performance Across Condition Categories.}
The a2z-1 model delivered solid performance across a variety of abdominal conditions, as shown by its AUC scores. For gastrointestinal conditions, the model achieved AUCs of 0.937 for appendicitis, 0.940 for diverticulitis, 0.855 for colitis, and 0.958 for small bowel obstruction. In vascular conditions, it showed AUCs of 0.876 for coronary artery calcification, 0.868 for aortic dissection, and 0.970 for unruptured aortic aneurysm. The model also performed well for hepatobiliary and pancreatic conditions, with scores of 0.940 for cholecystitis, 0.914 for biliary ductal dilatation, 0.890 for hepatic steatosis, 0.966 for liver cirrhosis, and 0.961 for acute pancreatitis (Figure \ref{fig:pancreatitis}). For renal conditions, the model showed AUCs of 0.949 for obstructive kidney stones, 0.873 for pyelonephritis, and 0.956 for hydronephrosis. Other abdominal conditions saw AUCs of 0.912 for hiatal hernia, 0.939 for splenomegaly, 0.960 for retroperitoneal hemorrhage, 0.921 for peritoneal free air (Figure \ref{fig:free-air}), 0.920 for peritoneal abscess, and 0.891 for pneumonia. Figure \ref{fig:radial-plot} offers a visual summary of a2z-1’s performance across the evaluated conditions.

\begin{figure*}[t!]
    \centering
    \includegraphics[width=\linewidth]{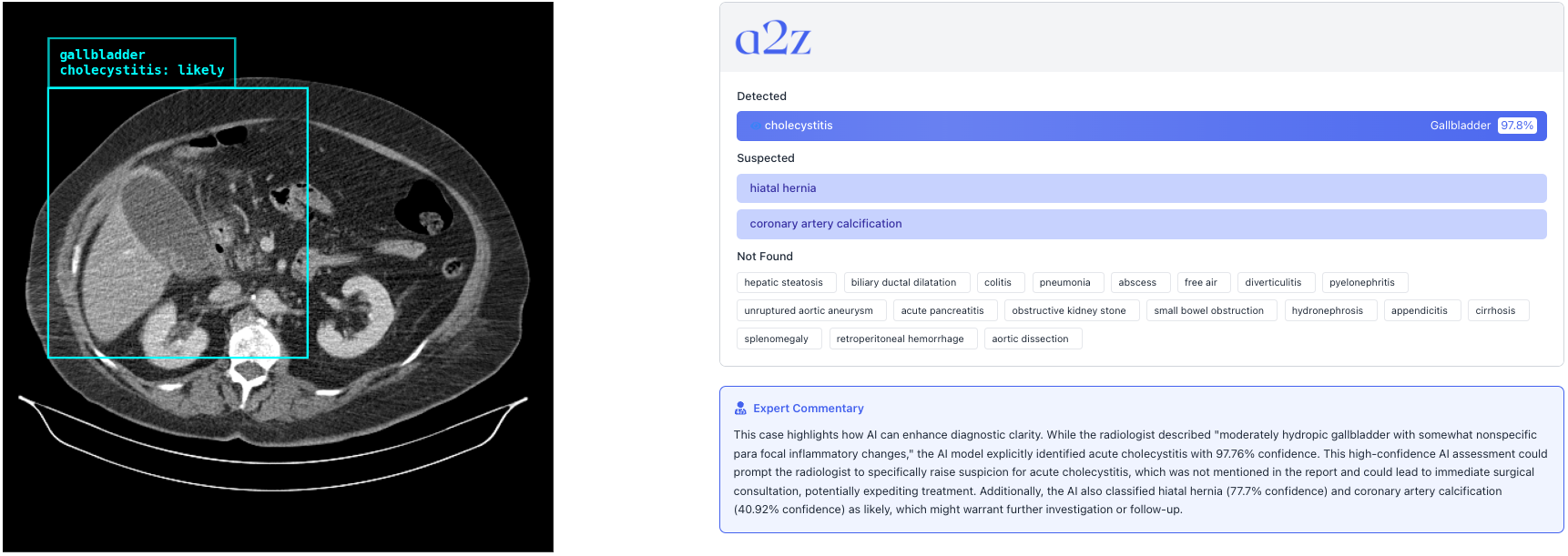}
    \caption{\textbf{a2z-1 enhances diagnostic confidence in acute cholecystitis case.}}
    \label{fig:cholecystitis}
\end{figure*}

\paragraph{Evaluation Details.} The model's effectiveness was evaluated through external validation, utilizing data from two independent health systems that did not contribute to the model's training. This external validation dataset consisted of 5444 studies from 4907 patients, providing a reliable sample size to assess the model's performance in varied real-world clinical environments. The ground truth for these cases was established through automated label extraction from the corresponding clinical radiology reports. Automated methods for label extraction from radiology reports have advanced significantly in recent years and are becoming highly accurate. We manually validated the quality of this ground truth by comparing the extracted labels to the actual content of the reports over a sample of reports that included every condition, achieving an average accuracy across findings of 99.4\% in this subset, with a F1 score of 92.6\%, recall of 99.05\% and precision of 88.7\%,  ensuring a high level of fidelity. All reports were signed by US-certified radiologists, further ensuring the reliability of the ground truth used for model evaluation.

\begin{figure*}[t!]
    \begin{center}
    \includegraphics[width=0.8\linewidth]{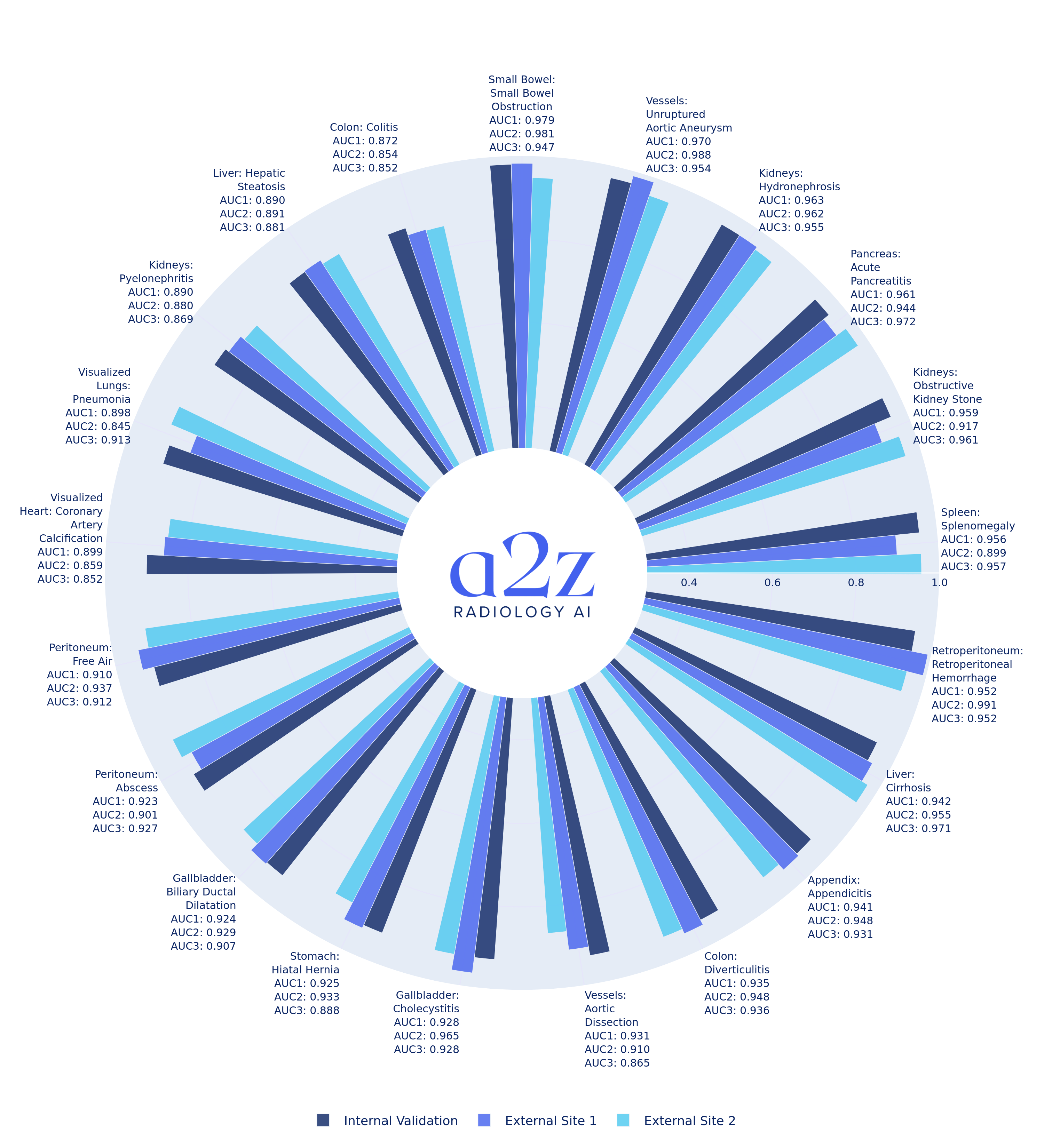}
    \end{center}
    \caption{ \textbf{a2z-1 model performance across internal and external validation sets.} The model shows consistent AUCs across sites, particularly for small bowel obstruction and appendicitis. Improved performance for liver cirrhosis and retroperitoneal hemorrhage in external sites suggests differences in patient populations. While slight variability is seen in coronary artery calcification and aortic dissection, overall performance remains strong across clinical environments.}
    \label{fig:auc-performance}
\end{figure*}

\subsection{Generalizability Across Radiology Sites/Practices}

\paragraph{Internal and External Validation.}
A key strength of a2z-1 is its ability to maintain high performance across different clinical sites, demonstrating robust generalizability. In this evaluation, \textit{internal validation} refers to testing the model on a separate dataset that was held out during training but comes from the same data source as the training set. This provides a baseline to assess the model’s performance within a familiar clinical environment. In contrast, \textit{external validation} refers to testing the model on independent datasets from two different states, which were not used during model development. These external validation sites allow us to assess the model's ability to generalize to distinct clinical settings and patient populations, ensuring real-world applicability.

The internal validation dataset, consisting of 9223 studies from 7234 patients, was used during model development to select the best-performing model. Importantly, the model was trained on a completely separate dataset with no patient overlap, ensuring that the internal validation results are unbiased. Meanwhile, the two external validation datasets come from two states different from the one used in model development, further testing the model’s generalizability across diverse geographic locations.

\begin{figure*}[t!]
    \centering
    \includegraphics[width=\textwidth]{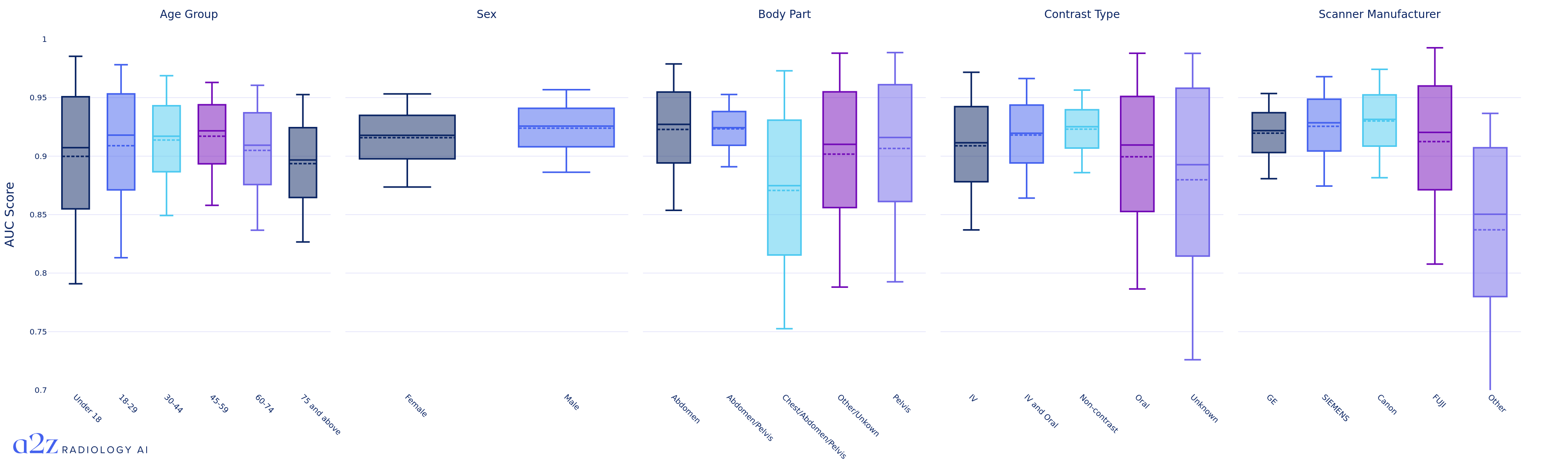}
    \caption{\textbf{AUC scores by demographic and imaging subgroups.} The model maintains consistent performance across different age groups, sexes, scanner manufacturers, and body parts. There's a slight decrease in performance for the youngest (Under 18) and oldest (75 and above) age groups, possibly due to anatomical differences or disease prevalence in these populations.}
    \label{fig:auc-scores-subgroup}
\end{figure*}

\paragraph{Consistent Performance Across Sites.}
As shown in Figure \ref{fig:auc-performance}, the model demonstrates consistent performance across internal and external validation sets. For example, in detecting small bowel obstruction, a2z-1 achieves AUC scores of 0.979, 0.981, and 0.947 across the internal and the two external sites, respectively, showing only slight variation. Similarly, the detection of appendicitis is highly consistent, with AUCs of 0.941, 0.948, and 0.931 across the three validation sets, indicating reliable performance across different locations.

\paragraph{Variations and Improved Performance.}
Interestingly, the model performs even better on some conditions in external validation, suggesting that variations in patient populations or case difficulty could have contributed to the improved results. For example, in detecting liver cirrhosis, the AUC increases from 0.942 in the internal validation set to 0.955 and 0.971 in the external sites, indicating potentially more identifiable characteristics of cirrhosis in these external patient populations. Similarly, the model shows an increase in performance for retroperitoneal hemorrhage, where the AUC jumps from 0.952 internally to 0.991 at one external site, suggesting a difference in case difficulty or clinical presentation between the sites.

In contrast, some conditions exhibit more variability. For instance, in detecting coronary artery calcification, the model’s AUC ranges from 0.899 in internal validation to 0.859 and 0.852 at the external sites. Aortic dissection follows a similar trend, with AUCs ranging from 0.931 internally to 0.910 and 0.865 externally. Despite these variations, the model still performs well across all settings, indicating adaptability to different clinical practices and patient populations.

\subsection{Consistent Performance Across Patient Characteristics and Imaging Protocols}

\paragraph{Analysis Overview.}
To evaluate the generalizability of the a2z-1 model, we conducted a comprehensive analysis using external validation data. The analysis focused on various patient demographics and imaging protocols, with relevant metadata extracted directly from DICOM files and associated radiology reports. These metadata fields, including patient age, sex, scan areas, contrast type used, and scanner manufacturer, allowed for a thorough assessment of model performance across critical subgroups. Table~\ref{tab:dataset-summary} shows the distribution of samples across different categories.

\begin{table}[htbp]
\centering
\caption{External validation dataset summary.}
\label{tab:dataset-summary}
\small
\renewcommand{\arraystretch}{0.9}
\setlength{\tabcolsep}{4pt}
\begin{tabular}{>{\raggedright\arraybackslash}p{0.7\columnwidth}>{\raggedleft\arraybackslash}p{0.2\columnwidth}}
Category & Count \\
\hline
\noalign{\smallskip}
Total & 5444 \\
\noalign{\smallskip}
\multicolumn{2}{l}{\textbf{Age Group}} \\
Under 18 & 259 \\
18-29 & 624 \\
30-44 & 1003 \\
45-59 & 1343 \\
60-74 & 1168 \\
75 and above & 1044 \\
Not specified & 3 \\
\noalign{\smallskip}
\multicolumn{2}{l}{\textbf{Sex}} \\
Female & 3078 \\
Male & 2364 \\
Other & 1 \\
Unknown & 1 \\
\noalign{\smallskip}
\multicolumn{2}{l}{\textbf{Body Part}} \\
Abdomen/Pelvis & 4309 \\
Abdomen & 605 \\
Chest/Abdomen/Pelvis & 240 \\
Pelvis & 84 \\
Other/Unknown & 206 \\
\noalign{\smallskip}
\multicolumn{2}{l}{\textbf{Contrast Type}} \\
No contrast & 2652 \\
IV and Oral & 1569 \\
IV only & 955 \\
Oral only & 180 \\
Unknown & 88 \\
\noalign{\smallskip}
\multicolumn{2}{l}{\textbf{Slice Thickness (mm)}} \\
$<$ 3 & 1704 \\
3-5 & 343 \\
$\geq$ 5 & 3397 \\
\noalign{\smallskip}
\multicolumn{2}{l}{\textbf{Manufacturer}} \\
GE & 2523 \\
Siemens & 1861 \\
Canon & 898 \\
Fuji & 111 \\
Other & 51 \\
\end{tabular}
\end{table}

\paragraph{Performance Across Age and Sex.}
The results indicate that a2z-1 demonstrates strong consistency in performance across the different subgroups, even in the external validation setting. Across age groups, the model shows high performance, with AUC scores ranging from approximately 0.89 for patients 75 and above to 0.92 for those aged 45-59. Although there is a slight dip in performance for the youngest (under 18) and oldest (75 and above) populations, this may be due to anatomical differences or lower disease prevalence in these groups. Despite these small variations, the model remains highly effective across the majority of age categories, with performance particularly strong in early and middle-aged adult groups (18-59), where AUC scores consistently exceed 0.91. Sex-based analysis shows minimal variation in performance, with AUC scores of ~0.92 for both males and females.

\paragraph{Performance Across Scan Areas.}
The model also performs well across different scan areas. For abdominal scans, the AUC hovers around 0.92, while pelvic scans show a slightly lower performance at 0.91. However, the model exhibits a minor dip when evaluating more complex multi-regional scans, such as those covering the chest, abdomen, and pelvis, where performance drops to around 0.87. This suggests that multi-regional imaging may introduce additional complexity, but the model still maintains a robust level of accuracy in these cases.

\begin{figure*}[t!]
    \centering
    \includegraphics[width=\linewidth]{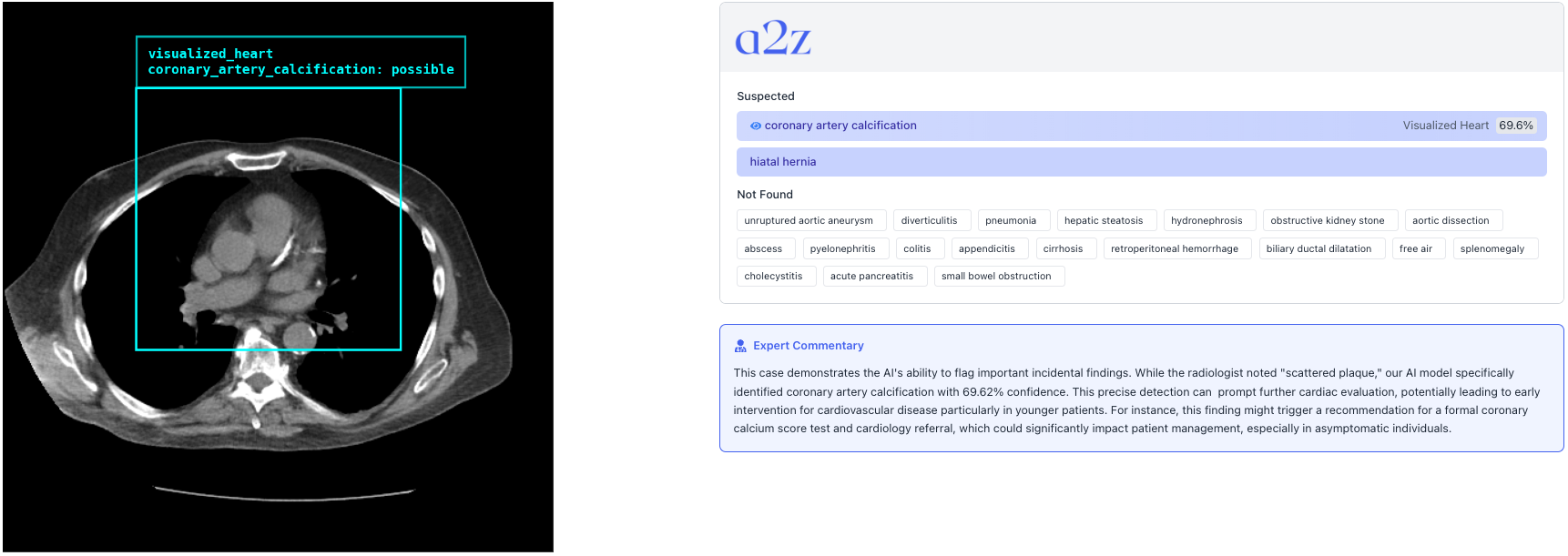}
    \caption{\textbf{a2z-1 flags incidental coronary artery calcification.}}
    \label{fig:cac}
\end{figure*}

\paragraph{Performance Across Imaging Protocols and Equipment.}
In terms of imaging protocols, a2z-1 adapts effectively across different contrast types. For IV contrast scans, the model achieves an AUC of around 0.91, and non-contrast scans see a comparable performance, with AUCs averaging around 0.92. When examining the performance across scanner manufacturers, the model demonstrates adaptability. It maintains AUCs above 0.92 for scans acquired on GE, Siemens, and Canon scanners. There is more variability when dealing with lesser-known or unspecified manufacturers, but the model remains reliable overall, with a narrow performance range across different equipment types.

\subsection{Confidence-Based Categorization for Streamlined Workflow Management}

\paragraph{Confidence Levels and Threshold Setting.}
The a2z-1 model employs a confidence-based three-tiered categorization system—"likely," "possible," and "unlikely"—to balance precision and recall. The thresholds for each category were set using internal validation data, with the "likely" category targeting a precision of 0.8, and the "possible" category targeting a precision of 0.4 (after excluding cases already marked as "likely"). These thresholds were designed to ensure that the model prioritizes high-confidence cases for immediate attention while capturing additional cases with lower certainty for further review. This three-tiered system allows for better handling of potential findings without overwhelming the workflow, minimizing "alarm fatigue."

\begin{figure*}[ht]
    \centering
    \includegraphics[width=\textwidth]{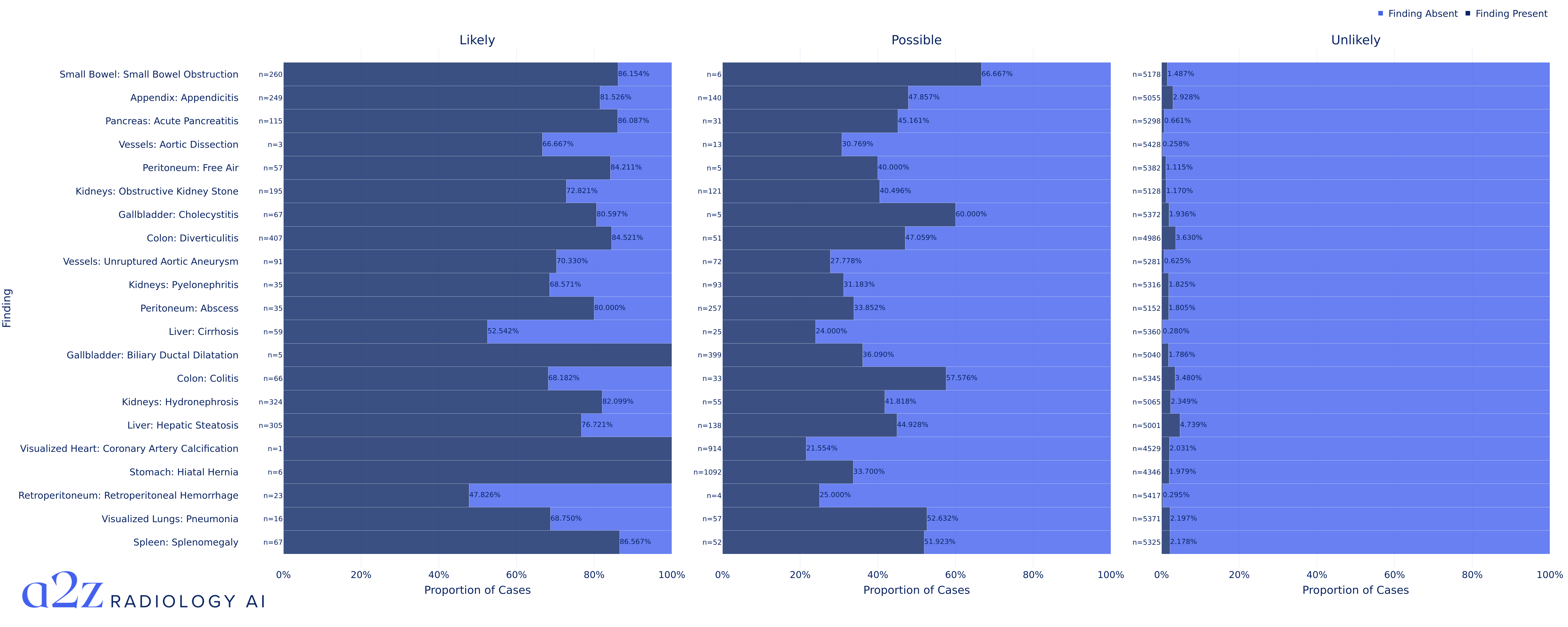}
    \caption{\textbf{Prediction accuracy across likelihood categories for various pathologies.} The model shows high accuracy in detecting "likely" and "unlikely" cases for most pathologies, with more variability in "possible" cases. This categorization enables flexible thresholding for various clinical applications.}
    \label{fig:prediction-accuracy}
\end{figure*}

\paragraph{High-Confidence Findings in the Likely Category.}
The "likely" category, designed to capture high-confidence findings with a precision threshold of 0.8, ensures that the most critical cases are flagged accurately. In the external validation dataset, the model identified 224 out of 305 small bowel obstruction cases (73.44\%) as "likely." For appendicitis, 203 out of 418 cases (48.56\%) were classified as "likely," while 99 out of 148 cases (66.89\%) for acute pancreatitis were captured in the "likely" category. This high-confidence filtering allows workflows to focus on cases that require prompt action without sacrificing accuracy.

\paragraph{Broader Coverage with Reduced Workflow Disruption.}
The "possible" category, with a precision threshold of 0.4 among cases not categorized as "likely," is designed to capture cases that are not entirely certain but may require more careful review. This classification helps identify findings that warrant additional scrutiny without necessarily being clear-cut positive cases. For appendicitis, 31.16\% of remaining cases were categorized as "possible," broadening the scope for further investigation. This tiered confidence system ensures that workflows are not overwhelmed while maintaining broad coverage for potential findings that may require attention.

\subsection{Analysis of High-Confidence Model Predictions in External Validation}

This analysis focuses on cases from the external validation dataset where the a2z-1 model predicted a pathology with high confidence (categorized as "likely") but the corresponding ground truth labels, derived from radiology reports, indicated the pathology was absent. The goal was to determine whether these divergences were true false positives or if errors in the report-to-label process or clinical ambiguity contributed to the discrepancies.

\paragraph{Labeling Errors in Certain Cases.}
In some instances, the model’s high-confidence predictions were correct, but the corresponding labels were incorrectly marked as negative due to errors in the report-to-label conversion process. For example, in one case, the model confidently predicted the presence of retroperitoneal hemorrhage. The report noted, "subcapsular and perinephric hematoma," consistent with the model's prediction, but the ground truth label incorrectly indicated that no retroperitoneal hematoma was present. Similarly, in another case, the model accurately predicted an obstructive kidney stone, and the report explicitly stated, "a 2 mm stone in the distal left ureter causing mild hydronephrosis," yet the label was marked negative.  While the proportion of labeling errors were less than 1\%, these were overrepresented in the model error analysis, suggesting that a portion of the model's false positives stemmed from inaccuracies in the labeling process rather than issues with the model's performance. This overrepresentation doesn't indicate a systemic problem with the labeling methodology, but rather highlights the challenges in accurately labeling edge cases or ambiguous instances where the report is not completely explicit.

\paragraph{Detection of Similar Pathologies.}
In several instances, we observed that the model identified pathological findings that, while not explicitly mentioned in the radiological report, were similar to the actual conditions and carried similar clinical implications. This phenomenon suggests that the model may be capturing underlying patterns indicative of broader disease categories rather than specific diagnoses. 
For example, in one case, the model predicted colitis in a patient who was diagnosed with diverticulitis of the descending colon. Both conditions represent inflammatory processes affecting the colon, albeit with distinct etiologies and manifestations. This misclassification, while not entirely accurate, demonstrates the model's ability to detect inflammatory changes in the colonic region.
Another illustrative case involved a patient with radiological findings consistent with hepatic steatosis and portal hypertension. The report detailed fatty infiltration of the liver, accompanied by hepatosplenomegaly and recannulized paraumbilical veins. The model, however, predicted cirrhosis. While not precisely correct, this prediction aligns with the constellation of findings associated with chronic liver disease, including fatty infiltration, which can be a precursor to cirrhosis, and signs of portal hypertension such as recannulized paraumbilical veins.
These examples underscore that in some false positive cases, the model identified clinically relevant abnormalities or conditions within the same pathological spectrum. This suggests that the model's performance extends beyond simple binary classification, demonstrating an ability to recognize complex pathological patterns. Importantly, these findings underscore the potential utility of the model in clinical practice, where its ability to flag abnormalities—even if not perfectly classified—could serve as a valuable tool for radiologists, directing their attention to areas of concern that warrant closer examination. In cases such as the diverticulitis misclassified as colitis, the model's identification of an abnormality in the correct anatomical region could still prompt a thorough evaluation, potentially expediting the diagnostic process and enhancing overall radiological assessment accuracy. However, they also highlights the need for nuanced evaluation metrics that can capture these predictions demonstrating accurate categorization of broad pathological classes despite imprecise identification of specific disease entities, and for careful interpretation of the model's outputs in conjunction with expert clinical judgment.

\begin{figure*}[t!]
    \centering
    \includegraphics[width=\linewidth]{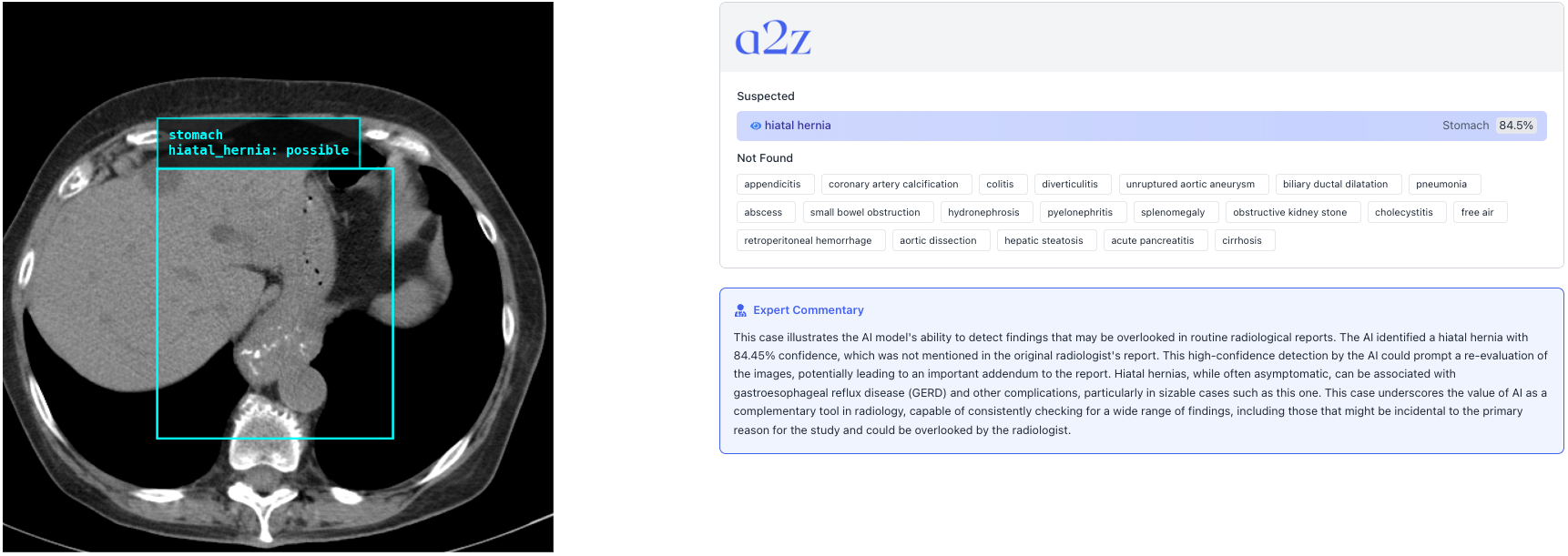}
    \caption{\textbf{a2z-1 detects hiatal hernia overlooked in routine radiology report.}}
    \label{fig:hiatal-hernia}
\end{figure*}

\paragraph{Manual Review of False Positives.}
A manual review of select high-confidence false positive cases uncovered additional insights. In one instance, the model predicted acute pancreatitis with high confidence, and while the report stated, "the pancreas is unremarkable," further review revealed subtle swelling of the pancreatic head and body with some surrounding stranding and fluid, consistent with acute pancreatitis (Figure \ref{fig:pancreatitis}). This suggests the model may have correctly identified the condition, which was not recognized in the original interpretation. In another case, the model confidently predicted cholecystitis, while the radiologist noted "questionable mild pericholecystic inflammatory changes" without explicitly raising the possibility of cholecystitis (Figure \ref{fig:cholecystitis}). This example highlights the AI's potential to reduce diagnostic uncertainty and minimize hedging in radiology reports. This capability not only aids in decision-making for borderline cases but also promotes more actionable reports, potentially leading to more timely and appropriate patient management. Ultimately, this synergy between AI and human expertise can enhance diagnostic clarity and improve the overall quality of radiological interpretations. Our analysis revealed additional instances where our model identified non-urgent incidental findings, such as coronary artery calcification (Figure \ref{fig:cac}) and hiatal hernias (Figure \ref{fig:hiatal-hernia}), which were present in the imaging studies but not documented in the corresponding radiological reports. This finding underscores the potential utility of the a2z-1 model in augmenting the consistency and comprehensiveness of radiology reporting, particularly for incidental findings that may have clinical relevance but are not the primary focus of the examination.

\subsection{Analysis of High-Confidence Model Misses in External Validation}

This analysis focuses on cases from the external validation dataset where the a2z-1 model predicted a low probability for a pathology that was labeled as positive based on the radiology report. The goal was to understand why the model failed to detect these findings and what contributed to the divergences.

\paragraph{Subtle and Mild Cases.}
Analysis of false negative cases revealed that approximately 40\% involved subtle or mild manifestations of the pathologies in question. In these instances, the radiological reports often characterized the findings as mild or early-stage, or expressed uncertainty about their presence, using phrases such as "cannot exclude appendicitis." This pattern suggests that the majority of a2z-1 model's false negatives occur in cases where the pathological features are less pronounced and potentially subject to inter-observer variability even among experienced radiologists. While these early or mild presentations may generally carry lower clinical urgency compared to more advanced cases, they nonetheless represent an important area for potential improvement in our model.

\paragraph{Labeling Errors}
While less frequently than in false positive cases, we also identified labeling errors within false negative cases, including cases where biliary ductal dilatation or coronary artery calcification were labeled as present, but the report did not mention these findings.

\begin{figure*}[t!]
    \centering
    \includegraphics[width=\linewidth]{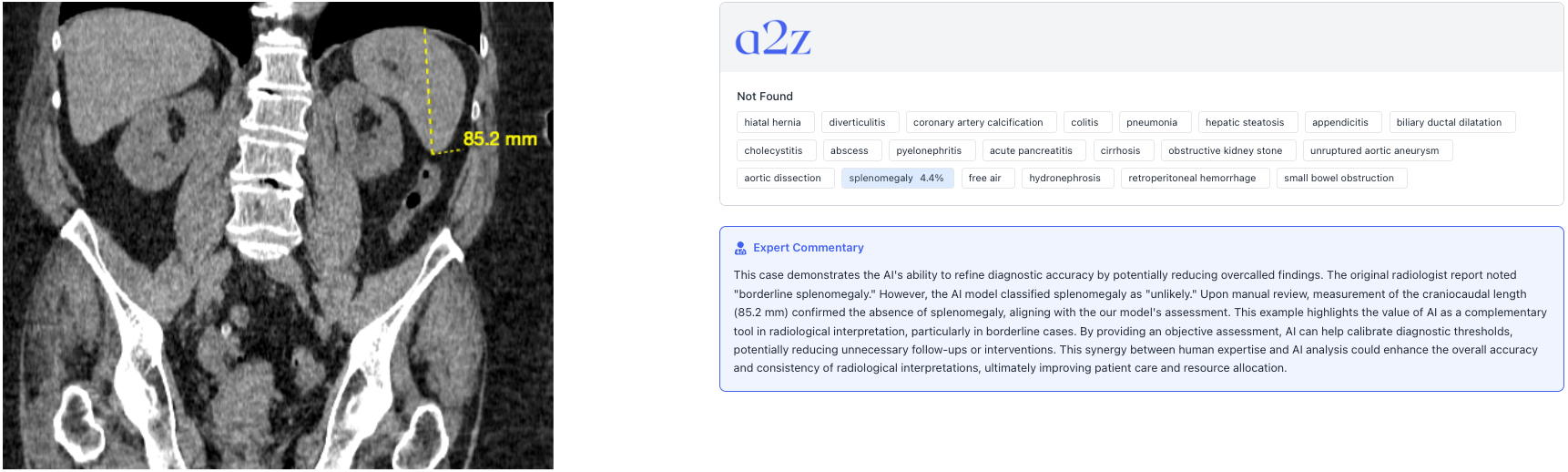}
    \caption{\textbf{a2z-1 refines diagnosis by correctly identifying absence of splenomegaly.}}
    \label{fig:splenomegaly}
\end{figure*}

\paragraph{Overcalled Incidental Findings}
In some other cases, we found that the radiology report mentioned incidental findings such as "borderline splenomegaly," which were not predicted by our model. However, on manual review, the spleen craniocaudal length was around 85 mm, which is below the threshold for diagnosing splenomegaly (Figure \ref{fig:splenomegaly}).

\section{Discussion}

\paragraph{Towards World-Class AI For Medical Image Interpretation.}
To the best of our knowledge, a2z-1 is the first AI model to demonstrate excellent performance across a broad spectrum of abdomen-pelvis CT pathologies validated both on internal and external sites. This breadth of application, covering 21 actionable and time-sensitive conditions, sets it apart from most existing models, which tend to focus on narrower use cases. Our external validation, carried out across multiple health systems, demonstrates the robustness of a2z-1 and its ability to generalize across different datasets and patient populations.

Unlike many existing studies that report results only on internal datasets, we conducted rigorous testing on independent external datasets to ensure the model's generalizability. The external validation shows that a2z-1 maintains its performance across different workflows, a key aspect for any model aiming to be broadly applicable. In particular, we achieve an average Area Under the Curve (AUC) of 0.92 across the 21 conditions. This performance level is significant because AUC values of this range typically reflect strong discriminative ability, ensuring reliable detection of both positive and negative cases.

\paragraph{Strengths Compared to Existing AI Models.}
While AI models for chest X-ray (CXR) interpretation have made significant strides in covering a broad range of conditions \citep{Jones2021-hh}, the interpretation of CT scans introduces an additional layer of complexity. CT images require the simultaneous analysis of multiple anatomical regions and structures, demanding a more intricate approach for AI-driven solutions \citep{schmidt_ai-based_2024}. AI models have shown some early success \citep{vanderbecq_deep_2024, rajpurkar_appendixnet_2020, hata_deep_2021, brejnebol_artificial_2022}, and generalist or foundation models are beginning to emerge, but these approaches have not yet demonstrated the rigorous performance necessary for comprehensive CT detection tasks, particularly in abdomen-pelvis imaging.

Our focus with a2z-1 is on a structured, task-based evaluation of predefined, clinically actionable conditions. This deliberate focus ensures that the model is not only broad in scope but also precise in identifying time-sensitive pathologies. Importantly, we emphasize external validation across multiple sites, an approach that distinguishes our work from emerging generalist models that, while promising, have not yet demonstrated strong, consistent results in task-specific evaluations for CT scans.

Another example of a comprehensive approach in abdomen-pelvis CT is Merlin, a foundation model designed for broad abdominal imaging tasks \citep{blankemeier2024merlinvisionlanguagefoundation}. While Merlin represents an early effort in this domain, its external validation remains limited, and its reported performance metrics are not directly comparable to those of a2z-1. By contrast, a2z-1 has been rigorously validated across multiple institutions, with condition-specific AUCs that provide detailed insights into its real-world applicability.

\paragraph{Broader Workflow Enablement.}
While the model is highly relevant for radiologists, its utility is not limited to traditional radiology workflows. a2z-1’s outputs could enable a range of applications, such as assisting emergency teams in triage scenarios, supporting quality control processes, selection of cases for peer review, or enabling automated alerts in high-volume settings like inpatient wards. The model’s tiered confidence-based system—categorizing findings into "likely," "possible," and "unlikely"—offers flexibility, allowing different teams within the workflow to prioritize cases based on their specific needs.

For instance, high-confidence findings might be routed to specialists or flagged for immediate intervention, while lower-confidence findings could be used in secondary review processes. This flexible prioritization ensures that the model adapts to various workflows, whether in emergency departments, or quality assurance roles.

\paragraph{Commercial Tools and Broader Applications.}
In the U.S., the only commercial AI tools approved for abdomen-pelvis CT interpretation are limited to three specific applications: BriefCase-Aortic Dissection Triage, BriefCase-Intra-abdominal Free Gas Triage, and Viz AAA. These tools are focused on triaging specific acute conditions, particularly in emergency settings, but they do not offer the breadth of support needed for a wide range of abdominal pathologies.

The scope of a multi-disease model like a2z-1 opens up the possibility for far more versatile workflows. With its ability to detect a wide range of conditions, a model like this could streamline processes, impacting areas such as patient triage, early detection of complications, or even monitoring patients during treatment courses. These broader applications could reduce time-to-action and potentially improve outcomes in high-stress, high-volume environments, where rapid decision-making is key.

\paragraph{Future Validation and Collaboration Opportunities.}
While our external validation results demonstrate strong performance across multiple health systems, we recognize the importance of continued validation across diverse clinical settings and patient populations. To facilitate this, we invite healthcare institutions to participate in further validation studies through our design partner program, available at \href{https://a2zradiology.ai/}{a2z Radiology AI}. This program offers opportunities for institutions to evaluate a2z-1's performance in their specific clinical environment while contributing to the broader understanding of AI's role in abdomen-pelvis CT interpretation. Partners can assess the model's integration into existing workflows, validate performance across their patient populations, and participate in collaborative research efforts. This approach not only enables robust external validation but also helps ensure that future development aligns with real-world clinical needs across diverse healthcare settings.

\bibliography{main.bib}
\end{document}